\documentclass[lineno]{jfm}

\usepackage{graphicx}
\usepackage{newtxtext}
\usepackage{newtxmath}
\usepackage{mathtools}
\usepackage{stmaryrd}
\usepackage{natbib}
\usepackage{hyperref}
\hypersetup{
    colorlinks = true,
    urlcolor   = blue,
    citecolor  = blue,
    linkcolor = blue,
}
\hypersetup{breaklinks=true}

\newcommand{\RomanNumeralCaps}[1]
\linenumbers

\title{Stokes flow of an evolving fluid film with arbitrary shape and topology}
\author{Cuncheng Zhu\footnote{Department of Mechanical and Aerospace Engineering, University of California San Diego, USA, cuzhu@ucsd.edu.}, David Saintillan\footnote{Department of Mechanical and Aerospace Engineering, University of California San Diego, USA, dstn@ucsd.edu.}, Albert Chern\footnote{Department of Computer Science and Engineering, University of California San Diego, USA, alchern@ucsd.edu.}}

\def\BDelta{\boldsymbol{\Delta}}
\def\Bnabla{\boldsymbol{\nabla}}

\def\Bvarphi{\boldsymbol{\varphi}}

\def\tA{\mathsfbi {A}}

\def\tE{\mathsfbi {E}}
\def\tF{\mathsfbi {F}}

\def\tH{\mathsfbi {H}}
\def\tI{\mathsfbi {I}}

\def\tL{\mathsfbi {L}}

\def\tQ{\mathsfbi {Q}}

\def\tS{\mathsfbi {S}}
\def\tT{\mathsfbi {T}}

\def\tg{\mathsfbi {g}}

\def\tt{\mathsfbi {t}}

\def\bB{\boldsymbol{B}}

\def\bF{\boldsymbol{F}}

\def\bN{\boldsymbol{N}}

\def\bU{\boldsymbol{U}}
\def\bV{\boldsymbol{V}}
\def\bW{\boldsymbol{W}}
\def\bX{\boldsymbol{X}}

\def\bb{\boldsymbol{b}}

\def\be{\boldsymbol{e}}

\def\bu{\boldsymbol{u}}
\def\bv{\boldsymbol{v}}
\def\bw{\boldsymbol{w}}

\def\Real{\mathbb{R}}
\def\First{\tI}
\def\Second{\tI \tI}

\def\GRAD{\operatorname{GRAD}}
\def\grad{\operatorname{grad}}
\def\div{\operatorname{div}}
\def\DIV{\operatorname{DIV}}

\DeclareMathOperator{\tr}{tr}

\DeclareMathOperator{\symm}{symm}

\newcommand{\thinsim}{{\raise.17ex\hbox{\(\scriptstyle\mathtt{\sim}\)}}}

\makeatletter
\newsavebox{\@brx}
\newcommand{\llangle}[1][]{\savebox{\@brx}{\(\m@th{#1\langle}\)}%
  \mathopen{\copy\@brx\kern-0.5\wd\@brx\usebox{\@brx}}}
\newcommand{\rrangle}[1][]{\savebox{\@brx}{\(\m@th{#1\rangle}\)}%
  \mathclose{\copy\@brx\kern-0.5\wd\@brx\usebox{\@brx}}}
\makeatother

\newcommand{\pair}[1]{\llbracket #1 \rrbracket}
\def\cf{cf.~}

\def\ie{i.e.,~}

\begin{document}
\maketitle

\begin{abstract}

The dynamics of evolving fluid films in the viscous Stokes limit is relevant to various applications, such as the modeling of lipid bilayers in cells. 
While the governing equations were formulated by Scriven in 1960, solving for the flow of a deformable viscous surface with arbitrary shape and topology has remained a challenge. 
In this study, we present a straightforward discrete model based on variational principles to address this long-standing problem.
We replace the classical equations, which are expressed with tensor calculus in local coordinates, with a simple coordinate-free, differential-geometric formulation.
The formulation provides a fundamental understanding of the underlying mechanics and directly translates to discretization.
We construct a discrete analogue of the system using the Onsager variational principle, which, in a smooth context, governs the flow of a viscous medium. 
In the discrete setting, instead of term-wise discretizing the coordinate-based Stokes equations, we construct a discrete Rayleighian for the system and derive the discrete Stokes equations via the variational principle. 
This approach results in a stable, structure-preserving variational integrator that solves the system on general manifolds.

\end{abstract}

\section{Introduction}
When surface dissipation dominates bulk dissipation (\ie when the Saffman–Delbrück length far exceeds the system size), the viscous flow of evolving fluid films can be modeled without accounting for the bulk fluid \citep{arroyo_relaxation_2009, saffman1975brownian}.
The governing equations, known as the evolving Stokes equations, have not only been of theoretical interest, but also found practical applications in engineering, such as the study of foam \citep{exerowa1997foam}. 
Their use also has a long-standing history in biophysical contexts, addressing fundamental problems across scales—from modeling subcellular structures like lipid bilayers to tissue-level phenomena such as epithelial monolayers \citep{al-izzi_active_2021}.

Historically, the governing equations were formulated by \citet{scriven1960dynamics} to study foam instability. 
When coupled with bending elasticity, the evolving Stokes equations serve as a common model for lipid membranes \citep{arroyo_relaxation_2009}. 
Recently, there has been substantial interest in modeling cell and tissue growth by coupling the viscous layer with additional anisotropy and active processes, such as those seen in the cellular cortex and active nematic fluids  \citep{al-izzi_morphodynamics_2022,  metselaar_topology_2019, torres-sanchez_modelling_2019}.
The evolving Stokes equations, along with extensions to the full Navier-Stokes equations, have been independently derived through various principles \citep{scriven1960dynamics, arroyo_relaxation_2009, torres-sanchez_modelling_2019, koba2017energetic, reuther_numerical_2020, sahu2020arbitrary, miura2018singular, al-izzi_morphodynamics_2022} and are compared in \citet{brandner2022derivations}. 
Our approach is closest to the formulation of \citet{arroyo_relaxation_2009} via the Onsager variational principle for dissipative systems. 

To computationally solve the evolving Stokes equations, most existing methods tackle the partial differential equations by explicitly splitting them into normal and tangential components. 
This involves a vector-valued equation for the tangent velocity and a scalar equation for the normal velocity on a manifold. 
When viewed this way, the system of equations is challenging to solve. 
First, the tangent equation is tensor-valued on an evolving Riemannian manifold, necessitating specialized techniques for covariant differentiation \citep{nestler2019finite, gross2018hydrodynamic, knoppel_globally_2013, torres2020approximation, voigt2019fluid}. 
Some methods avoid the tensor-valued equation of tangent velocity through streamfunction formulations, although this approach is limited to simply-connected domains \citep{torres-sanchez_modelling_2019} unless additional topological techniques are employed \citep{yin_cohomology}. 
Second, the equations in the decomposed form are explicitly coupled with surface curvatures, which, in classic numerical approximations, require representations using high-order basis functions \citep{sahu2020arbitrary, krause2023numerical}.

In this study, instead of directly tackling the classic equations, we return to the fundamental governing kinematics and principles of a viscous surface film.
We replace the system of equations expressed in tangent-normal splitting with a simple, coordinate-free differential-geometric formulation. 
We show that our resulting equations agree with earlier formulations by \citet{scriven1960dynamics} and \citet{arroyo_relaxation_2009} but in a more elegant form.
Our approach also directly carries over to the discrete setting.
The abstraction directly leads to a discretization for the strain rate tensor on discrete meshes.
We then construct a discrete Rayleighian for the system and derive the discrete Stokes equations via the Onsager variational principle, leading to a simple linear system.
We provide a self-contained exposition of both the continuous and discrete models, as well as the numerical methods used to solve the system of equations on arbitrary geometries and topologies.

\section{Theory}
We derive the equations of motion for an evolving surface driven by viscous Stokes flow using Onsager's variational principle.  
Specifically, a dissipation functional, called the Rayleighian, is assigned to each position and velocity state of a deformable surface. 
 The equations of motion are then obtained by finding the velocity that minimizes this functional.

We adopt a differential geometric approach, similar to \citet{marsden1994mathematical}, to describe continuum mechanics.
Tensors are expressed using fiber bundle notation for clarity. 
Readers unfamiliar with this terminology can refer to \S\ref{sec: notation} for a brief overview.
We first define relevant tensors and differential operators  in \S\ref{sec: differential geometry}, followed by the definition of the Rayleighian for the system in \S\ref{sec: onsager principle}. 
For simplicity in presenting our smooth theory, all functions and tensors are assumed to be of \(C^\infty\) class.
 
\subsection{Terminology and notation}\label{sec: notation}
On a manifold $M$, the \emph{tangent space} $T_pM$ is a vector space that provides a local linear approximation of the manifold at $p \in M$.
The disjoint union of all tangent spaces forms the \emph{tangent bundle} \(TM = \bigcup_p T_pM\), where each \(T_pM\) is also called the \emph{fiber} of the tangent bundle $TM$ at the \emph{base point} \(p\in M\) on its \emph{base manifold} $M$.
The dual bundle of $TM$ is the \emph{cotangent bundle}, $T^*M$, whose fibers are the dual spaces of the tangent spaces.
The $(m, n)$-typed tensor bundle is written as $ TM^{\otimes m, n} =  TM^{\otimes m} \otimes T^*M^{\otimes n}$.
A tensor field $\boldsymbol{\Psi}$ on $M$ is a \emph{section} of the tensor bundle, denoted as \(\boldsymbol{\Psi}\in\Gamma(TM^{\otimes m, n})\), which assigns a tensor \(\boldsymbol{\Psi}|_p\in T_pM^{\otimes m, n}\) to each point \(p\in M\).
When a basis section is chosen, its components, $\Psi_{i_1 \ldots i_n}^{j_1 \ldots j_m}$, have $m$ contravariant indices and $n$ covariant indices. 

For a map $\varphi: M \rightarrow W$, where $W$ is another manifold, the \emph{pullback bundle} \(T_{\varphi}W = \varphi^*TW\) is defined with $M$ as its base manifold, while the fiber at \(p\in M\), \((T_\varphi W)|_p = T_{\varphi(p)}W\), is the tangent space to $W$ at $\varphi(p) \in W$.

Note that tensors can be identified with linear maps; for example, $\boldsymbol{\Psi}|_p \in T_pM^{\otimes m, n} $ acts as a linear map $\boldsymbol{\Psi}|_p: T_pM \rightarrow T_pM^{\otimes m, n-1}$.

\subsection{Differential geometry of an evolving surface}\label{sec: differential geometry}
We describe a deformable surface as follows. 
Let \(M\) be a 2-dimensional closed manifold (\(\partial M = \varnothing\)) with arbitrary genus, representing the \emph{material/Lagrangian space}.  
Let \(W\) be a 3-dimensional Riemannian manifold representing the \emph{world/Eulerian space}. 
Let \(\tg\in \Gamma(T^*W\otimes_{\rm symm}T^*W)\) (\ie a symmetric (0, 2) tensor field) denote the Riemannian metric tensor on \(W\).%
The common setup is to set \(W = \Real^3\), the Euclidean 3-space, with \(\tg\llbracket \bV,\bW \rrbracket = \bV \bcdot \bW\) being the Euclidean inner product for any \(\bV,\bW\in \Real^3\).
However, our theory is not limited to this assumption.
The \emph{position} of a deformable surface is an embedding function \(\varphi\colon M\hookrightarrow W\) from the material space to the world space. 
A time-dependent deformable surface, or an evolving surface, is a time-dependent embedding \(\varphi(t) = \varphi_t\colon M\hookrightarrow W\).
The \emph{material velocity} of an evolving surface \(\varphi(t)\) is given by \(\dot\varphi_t \coloneqq \partial_t\varphi_t\), which is a \emph{pullback} vector field  \(\dot\varphi_t\in\Gamma(T_{\varphi_t}W)\) with its base point in the material frame $M$ (\cf \S \ref{sec: notation}). %
\footnote{
The velocity \(\dot\varphi_t|_p\) is evaluated at each point \(p\in M\) in the material domain, and its value \(\dot\varphi_t|_p\in T_{\varphi_t(p)}W\) is a tangent vector to the world space based at \(\varphi_t(p)\in W\).}

Each material velocity \(\dot\varphi_t\in \Gamma(T_{\varphi_t}W)\) can be extended to a \emph{spatial velocity} over $W$, denoted by \(\bU_t\in \Gamma(TW)\), such that \(\dot\varphi_t = \bU_t\circ\varphi_t\), or equivalently, $\dot \varphi_t|_p = \bU_t|_{\varphi_t (p)}$ for $\varphi_t(p) \in W$.
Note that this extension is not unique, as the assignment of \(\bU_t\) at locations away from the surface \(\varphi_t(M)\subset W\) is arbitrary. 

\subsubsection{Deformation gradient and Cauchy--Green tensor}
The differential of the positioning  \(\varphi\colon M\hookrightarrow W\) of a surface is denoted by \(\tF\coloneqq d\varphi\in \Gamma(T_{\varphi}W\otimes T^*M)\).
This differential is also known as the pushforward map, the tangent map, or the \emph{deformation gradient}.
At each point \(p\in M\), the two-point tensor \(\tF|_p\) can be identified with a linear map (\cf \S\ref{sec: notation}) \(\tF|_p = d\varphi|_p\colon T_pM\rightarrow T_{\varphi(p)}W\) %
\footnote{
At each point \(p\in M\), the tensor \(\tF|_p\in T_{\varphi(p)}W\otimes T_{p}^*M\) pairs with a vector in \(T_pM\) and returns a value in \(T_{\varphi(p)}W\).
Since it relates quantities in two different spaces $M$ and $W$, it is called a two-point tensor.
}
that realizes a material tangent vector as a 3D world vector, as expected for the deformation gradient.

The embedding \(\varphi\colon M\hookrightarrow W\) induces a Riemannian metric (the first fundamental form) \(\First\in\Gamma(T^*M\otimes_{\rm symm}T^*M)\) on \(M\) from the Riemannian structure \(\tg\) on \(W\).  
This induced metric is defined by \(\First\llbracket\bv,\bw\rrbracket\coloneqq\tg\llbracket \tF\bv,\tF\bw\rrbracket = \bv^\top\tF^\top\tg\tF\bw\) for each \(\bv,\bw\in T_pM\), \(p\in M\).  
In continuum mechanics, \(\First = \tF^\top\tg\tF\) is known as the \emph{right Cauchy--Green tensor}.%
\footnote{
Here we identify tensors \(\tI, \tg, \tF\) as linear maps \(\tI_p \colon T_pM \rightarrow T_p^*M\), \(\tg_q \colon T_qW \rightarrow T_q^*W\), \(\tF_p\colon T_pM\rightarrow T_{\varphi(p)}W\). 
Then we have the compositional relation \(\tI = \tF^\top \tg\tF\), where \(\tF_p^\top\colon T_{\varphi(p)}^*W\rightarrow T_p^*M\) is the adjoint of \(\tF_p\).
}

For brevity, we denote the induced metric as \(\langle\cdot,\cdot\rangle = \First\llbracket\cdot,\cdot\rrbracket\) and the \(L_2\) inner product as \(\llangle\cdot,\cdot\rrangle = \int_M\langle\cdot,\cdot\rangle\, \mathrm{d}A\) where \(\mathrm{d} A\) is the area form from the metric \(\First\).
These inner products define the vector norm \(|\cdot|^2\) and the \(L_2\) norm \(\|\cdot\|^2\).
Dual pairings between primal (contravariant) and dual (covariant) tensors are denoted by $\langle \cdot | \cdot \rangle$ and \(\llangle \cdot | \cdot \rrangle = \int_M \langle \cdot | \cdot \rangle \mathrm{d} A\).
The \(L_2\) inner product between two scalar functions $f$ and $g$ is denoted by \(\llangle f, g\rrangle = \int_{M}fg\, \mathrm{d} A\).

\subsubsection{Strain rate tensor}\label{sec: strain rate tensor}
For an evolving surface \(\varphi_t\), the induced metric \(\First_t\in \Gamma(T^*M\otimes_{\rm symm}T^*M)\) varies with time $t$.
The \emph{strain rate tensor} \(\tE_t\in \Gamma(T^*M\otimes_{\rm symm}T^*M)\) is defined as half the rate of change of the induced metric: \(2\tE_t \coloneqq \dot{\First_t}\).

To compute the time derivative of \(\First_t = \tF^\top_t\tg\tF_t\) we first calculate the time derivative of \(\tF_t\).  
Let \(\bU_t\in\Gamma(TW)\) be any extended velocity field such that \(\dot\varphi_t = \bU_t\circ\varphi_t\).  
Then,
\begin{align}\label{eqn: deformation gradient dot}
   \dot \tF 
   = d \dot \varphi_t =
   d (\bU_t \circ \varphi_t) = {\bnabla} (\bU_t \circ \varphi_t) \circ d \varphi_t \equiv (\bnabla \bU_t) \tF,
\end{align}
where \({\bnabla: \Gamma (TW) \rightarrow \Gamma(TW\otimes T^*W)}\) is the Levi-Civita covariant derivative compatible with the metric \(\tg\).
When applied to a vector \(\bv\in\Gamma(TM)\), \(\dot\tF_t\bv = {\bnabla}_{\tF_t\bv}\bU_t\), which is a directional derivative of \(\bU_t\) tangential to the surface.  
In particular, \eqref{eqn: deformation gradient dot} depends only on the values of \(\bU_t\circ\varphi_t\) at the surface and not on the choice of its extension \(\bU_t\) over \(W\).

With a time-invariant metric $\tg$ and \eqref{eqn: deformation gradient dot}, the strain rate tensor is given by
\begin{align}\label{eqn: strain rate}
    2\tE_t = \dot \First_t = \dot \tF_t^\top \tg \tF_t + \tF_t^\top \tg \dot \tF_t = \tF_t^\top (({\bnabla} \bU_t)^\top \tg + \tg {\bnabla} \bU_t ) \tF_t.
\end{align}
When applied to vectors, $ 2\tE_t \pair{\bv, \bw}|_p = ({\bnabla}_{\bF_t \bv} \bU_t)\bcdot(\bF_t \bw) + ({\bnabla}_{\bF_t  \bw} \bU_t)\bcdot(\bF_t  \bv)$ for $\bv, \bw \in \Gamma(T_pM)$.
Again, \(\tE_t\) depends only on the values of \(\bU_t\circ\varphi_t\) and not on the choice of its extension \(\bU_t\).

\subsubsection{Differential operators} \label{sec: differential operators for an evolving surface}
The operator that takes each velocity field \(\dot \varphi = \bU\circ\varphi\) at the surface to its induced strain rate tensor \eqref{eqn: strain rate} is called the Killing operator:
\begin{align}\label{eqn: evolving killing}
    \mathcal{K}: \Gamma(T_{\varphi}W) \rightarrow \Gamma(T^*M \otimes_{\operatorname{symm}} T^*M), \quad \mathcal{K} \dot \varphi \coloneqq \tE =  \tF^\top \frac{({\bnabla}\bU)^\top\tg + \tg{\bnabla}\bU}{2}\tF.
\end{align}
The trace of the strain rate tensor defines the \emph{divergence operator}
\begin{align}\label{eqn: evolving divergence}
    \DIV \colon \Gamma(T_\varphi W) \rightarrow C^{\infty}(M),\quad \operatorname{DIV} \dot \varphi \coloneqq \tr (\mathcal{K} \dot \varphi),
\end{align}
which measures the rate of change of area induced by the given velocity field. 
Define the \emph{gradient operator} as the negative adjoint of the divergence operator:
\begin{align}\label{eqn: evolving gradient}
    \GRAD = - \DIV^*: C^{\infty}(M) \rightarrow \Gamma(T_\varphi^*W),\quad \llangle\GRAD f|\dot \varphi\rrangle = - \llangle f, \DIV\dot \varphi\rrangle
\end{align}
for all \(f\in C^\infty(M)\).
The \emph{covariant divergence} is the negative adjoint of the Killing operator:
\begin{align}\label{eqn: evolving covariant divergence}
    \DIV^{\bnabla} = - \mathcal{K}^*: \Gamma(TM \otimes_{\symm} TM) \rightarrow \Gamma(T_\varphi^*W),\quad \llangle \DIV^{\bnabla} \tT|\dot \varphi\rrangle = -\llangle \tT|\mathcal{K}\dot \varphi\rrangle
\end{align}
for all symmetric tensor fields \(\tT\in\Gamma(TM\otimes_{\rm symm}TM)\).  
Here the dual pairing \(\langle \tT|\tE\rangle\) between \((0,2)\) and \((2,0)\) tensors is \(\langle \tT|\tE\rangle = \tr(\tT^\top\tE)= \sum_{ij} T^{ij}E_{ij}\).

Notably, the operators $\mathcal{K}$, $\DIV$, $\GRAD$, and $\DIV^{\bnabla}$, derived from the strain rate tensor $\tE$ of the evolving surface, differ from the commonly used intrinsic operators $\mathscr{K}$, $\div$, $\grad$, and $\div^{\bnabla}$.
As shown through the normal-tangent decomposition in Appendix~\ref{sec: explicit formulation}, these differences arise from additional contributions related to curvature (\cf \ref{eqn: killing normal-tangent}–\ref{eqn: covariant divergence normal-tangent}).

 \subsection{Onsager's variational principle for an evolving viscous fluid film } \label{sec: onsager principle}
The relaxation dynamics of an evolving viscous fluid film is governed by the Onsager's variational principle, which states that Stokes flow follows the dynamics of the least energy dissipation 
\citep{doi_onsagers_2011,arroyo_relaxation_2009}.
The dissipation functional, or Rayleighian, captures the system's dissipation rate, and the governing flow is the stationary condition of this functional subject to the incompressibility constraint.

Consider a viscous dynamical system where the dissipation functional $\mathcal{D}$ comprises the rate of viscous dissipation, and power inputs from an external force \(\bB\in\Gamma(T^*W)\) and a shape-dependent potential energy \(V_\varphi\):
\begin{align}
\begin{split}
    \mathcal{D}(\bU) &= \frac{1}{2} \llangle \tE | {\tT} \rrangle -\llangle \bB | \bU \rrangle + \dot V_\varphi \\
    & = \frac{1}{2} \int_M \langle \tE | \tT \rangle ~\mathrm{d} A  - \int_{\varphi(M)} \langle \bB | \bU \rangle ~\mathrm{d} A + \int_M \langle \delta_\varphi V_\varphi| \dot \varphi \rangle ~\mathrm{d}A.
\end{split}
\end{align}
Here, $\tT = 2 \mu \tE \in \Gamma(TM \otimes_{\operatorname{symm}} TM)$ is the viscous stress tensor (assumed Newtonian with constant viscosity $\mu$), $\delta_\varphi V_\varphi$ is the conservative force obtained by taking the variation of $V_\varphi$, and $\dot \varphi = \bU \circ \varphi$ is the material velocity. 

Onsager's variational principle determines the Stokes flow by solving the constrained optimization problem:
\begin{align}\label{eqn: Onsager}
\min_{\bU}~\mathcal{D}(\bU) \quad \text{subject to}
  \quad \DIV \bU = 0.
\end{align}
Using the fluid pressure $p \in C^\infty(M)$ as a Lagrange multiplier, this problem can be formulated as a minimax problem involving the augmented Rayleighian $\mathcal{R}$:
\begin{align}\label{eqn: Rayleighian}
\begin{split}
       \min_{\bU} \max_{p}~ \mathcal{R}(\bU, p), \quad \mathrm{where} \quad \mathcal{R}(\bU, p) = \mathcal{D}(\bU) + \llangle p , \DIV \bU \rrangle.
\end{split}
\end{align}
On a closed surface, the stationary conditions yield the Stokes equations for an evolving surface:
\begin{align}\label{eqn: stokes flow}
    2 \mu \DIV^{\bnabla} \mathcal{K} \bU - \GRAD p + \bB - \delta_{\varphi} V_\varphi &=  0,\qquad
    \DIV \bU = 0.
\end{align}
The symmetric, negative-definite, Laplacian-like operator $2\DIV^{\bnabla} \mathcal{K}: \Gamma(TW) \rightarrow \Gamma(T^*W) $ quantifies the viscous force, and is thus termed viscosity Laplacian.  
Although not the main focus of this paper, a common free energy in this context is the Helfrich Hamiltonian $V_{\varphi} = \int_W \kappa H^2 \mathrm{d} A$, which results in the bending force $\bB_\kappa = -\delta_\varphi V_\varphi =  \kappa \bN ( \Delta H +   2 H ( H^2 - K) )$, where $H$ and $K$ are the mean and Gaussian curvatures, respectively, $\kappa$ is the bending modulus, and \(\bN\) is the surface normal \citep{helfrich1973elastic}.
This bending-driven relaxation dynamics is also used in the numerical examples of \S\ref{sec: results}.
The commonly used form of \eqref{eqn: stokes flow}, expressed through normal-tangent decomposition, is derived in Appendix~\ref{sec: explicit formulation}.

\subsubsection{Remarks on the evolving Stokes equations and their solvability}\label{sec: solvability}

Note that the presence of a \emph{Killing vector field} \(\bX\) ($\mathcal{K} \bX = \mathsfbi{0}$) can make the solution to \eqref{eqn: stokes flow} non-unique.  
In that case we select the least-norm solution, effectively projecting out rigid motions.  
This treatment aligns with scenarios where the surface is immersed with friction in a bulk fluid. 
A simple way to model friction is by augmenting the Rayleighian to $\mathcal{R}^\alpha = \mathcal{R} + \alpha \| \bU - \bU_0 \|^2 / 2$, where $\bU_0$ is the bulk velocity and $\alpha$ is the friction coefficient. 
This introduces a friction force $\bB_\alpha = \alpha (\bU - \bU_0)$ in the momentum equation.
Such flow driven by the bulk fluid is used as a numerical example in \S\ref{sec: bulk flow}.
When $\bU_0 = \boldsymbol{0}$, the solution approaches the least-norm solution as $\alpha \rightarrow 0$.

\section{Methods}\label{sec: methods}
In this section, we present a simple numerical method to solve \eqref{eqn: stokes flow} on a triangle mesh with general geometry and topology. 
We begin by discretizing the strain rate tensor using finite elements, following the definition provided in \eqref{eqn: strain rate}.
Based on the discretized strain rate, we construct the system's Rayleighian.
A nonlinearly stable, structure-preserving variational integrator is derived based on the Onsager’s variational principle.
To keep the numerical scheme minimal, we demonstrate the method using first-order spatial and temporal discretizations.

\begin{figure}
\centerline{\includegraphics[width= 0.55\textwidth]{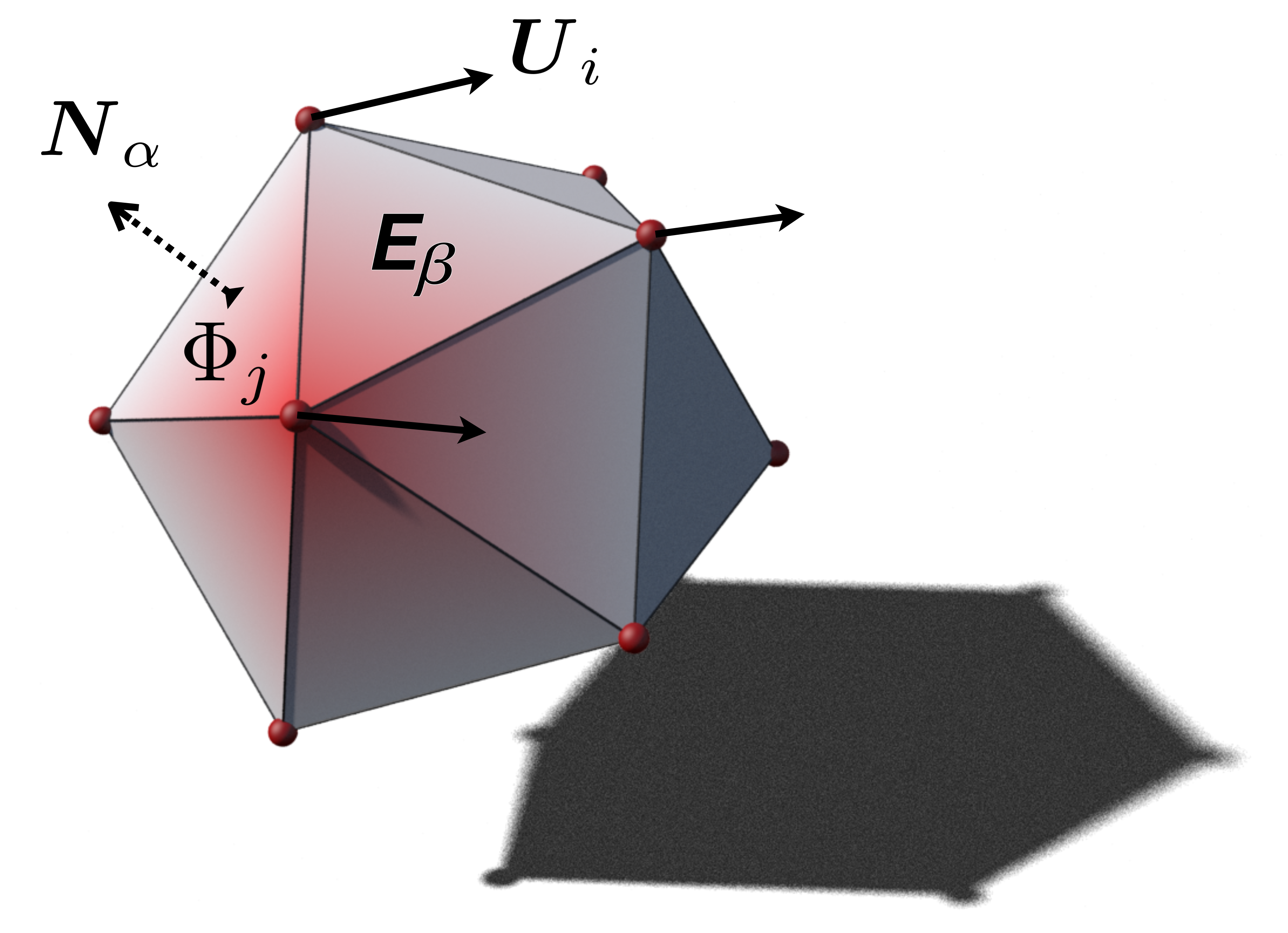}}%
  \caption{
  Discretization of the strain rate tensor. The velocity \(\bU_i\) is defined at each vertex $i$, with surface normals \(\bN_\alpha\) at each face $\alpha$. The finite-element hat function $\Phi_j$, shown by the color map, has local support around vertex $j$. Differentiating \(\bU_i\) via $\Phi_j$ produces $(\Bnabla \bU)_\beta$, which is symmetrized and projected using \(\mathcal{P}_\beta = \tg - \bN_\beta \otimes \bN_\beta \) to yield the strain rate tensor $\tE_\beta$ on each face $\beta$.}
\label{fig:scheme}
\end{figure}

We discretize the system on a closed triangular mesh $M$ of arbitrary genus, consisting of vertices, edges, and faces $\{\mathfrak{V, E, F}\}$.
The vertex positions are given by the realization $\Bvarphi: M \hookrightarrow \Real^3$. 
$M$ has a tangent space $T_\alpha M$ and a surface normal $\bN_\alpha$ at each face $\alpha \in \mathfrak F$ (\cf figure~\ref{fig:scheme}).

In the discrete setting, we express tensor-valued functions and operators using index notation, with upper indices denoting their components in Cartesian coordinates.
Vertex and face elements are denoted with lower indices, where vertex indices are Roman letters and face indices are Greek letters.
For example, we denote vertex positions as $\varphi_j^i$, face normals as $N_\alpha^i$, face pressure as $p_\alpha$, for $j \in \mathfrak V$, $\alpha \in \mathfrak F$, and $i \in \{1, 2, 3\}$.

\subsection{Strain rate tensor and differential operators}\label{sec: discrete strain rate tensor}
To express the strain rate tensor $\tE$ on $M$ (\cf \eqref{eqn: strain rate}) in Cartesian coordinate, we first realize it in $\Real^3$.
We denote this realization as $2 \tilde \tE = \mathcal{P}  (({\bnabla} \bU)^\top \tg + \tg {\bnabla} \bU ) \mathcal{P}$, which satisfies $\tilde \tE \pair {\bV, \bW} = \tE \pair {\mathcal{P} \bV, \mathcal{P} \bW}$ for $\bV, \bW \in \Gamma(\Real^3)$.
Here, the linear projector $\mathcal{P} = \tg - \bN \otimes \bN: \Gamma(\Real^3) \rightarrow \Gamma(TM)$ projects $\Real^3$ vectors onto the surface.
In the discrete setting (\cf figure~\ref{fig:scheme}), the strain rate tensor $\tilde E^{kj}_\alpha$ and Killing operator $\mathcal K_{\alpha l}^{kj i}$ are expressed in terms of the velocity $U_{j}^{i}$, projector $\mathcal{P}_{\alpha}^{ij} = \delta_{\alpha}^{ij} - N_{\alpha}^i N_{\alpha}^j$, and $\Real^3$ component-wise surface derivative $ ({\bnabla}_{\mathcal{P}})_{\alpha j}^i$ as:
\begin{align}\label{eqn: discrete strain rate tensor}
\begin{split}
   2 \tilde E^{kj}_\alpha = \sum_{l, i} 2 \mathcal{K}_{\alpha l}^{kj i} U_{l}^{i} 
    = \sum_{l, i} \left( \mathcal{P}_{\alpha}^{ik} ({\bnabla}_{\mathcal{P}})^j_{\alpha l}  + \mathcal{P}_{\alpha}^{ij} ({\bnabla}_{\mathcal{P}})^k_{\alpha l} \right)U_{l}^i. 
\end{split}
\end{align}
Component-wise, $({\bnabla}_{\mathcal{P}})_{\alpha j}^i = {{\bnabla}^i \Phi_{\alpha j} }$ follows scalar finite element discretization, where $\Phi_{\alpha j}$ is the hat function at vertex $j$ restricted to face $\alpha$.
The divergence \eqref{eqn: evolving divergence} is given by the trace $\DIV_{\alpha l}^i = \sum_j \mathcal{K}^{jj i}_{\alpha l}$.
The gradient \eqref{eqn: evolving gradient} and covariant divergence \eqref{eqn: evolving covariant divergence} can be obtained, up to a sign flip, by permuting the lower indices of the divergence and Killing operators, respectively.
\subsection{Variational integrator by Onsager's variational principle}\label{sec: discrete onsager variational principle}

Given the vertex position $\Bvarphi_{(0)}$ and pressure $p_{(0)}$ at the current time $t = 0$, with a time step $\epsilon > 0$, the velocity is discretized as $\bU = \mathring \Bvarphi / \epsilon =  (\Bvarphi - \Bvarphi_{(0)}) / \epsilon$, and the power input as $\dot V_{\Bvarphi} = (V_{\Bvarphi} - V_{\Bvarphi_{(0)}}) / \epsilon$.
State variables $\Bvarphi_{(\epsilon)}$ and $p_{(\epsilon)}$ at $t = \epsilon$ follow a time-incremental Onsager's variational principle on a discrete Rayleighian (\cf \S\ref{sec: onsager principle}), $\Bvarphi_{(\epsilon)}, p_{(\epsilon)} = \arg \min_{\Bvarphi} \max_p \mathcal{R}$, where
\begin{align}\label{eqn: discrete Rayleighian}
\begin{split}
    \mathcal{R}(\Bvarphi, p; \Bvarphi_{(0)}, p_{(0)}, \epsilon)
    &\coloneqq \epsilon^2 \left [ \mu \| \tilde \tE \|^2  + \llangle \DIV \bU, p \rrangle + \dot V_{\Bvarphi} - \llangle \bU| \bB \rrangle \right]\\
    &\equiv \mu \mathcal{K}_{\lambda m}^{kji}(A_{\lambda \gamma})^{-1} \mathcal{K}_{ \gamma n }^{kj l} \mathring \varphi|^i_m \mathring \varphi|_n^{l} + \epsilon ( \mathring \varphi|_{i}^l   \DIV_{\alpha i}^l p_{\alpha}   +  V_{\Bvarphi} -  \mathring \varphi|_{i}^l B_{i}^l    ) \\
    &\equiv \mu  \mathring \Bvarphi^\top \tL ~\mathring \Bvarphi  +  \epsilon \left [  (\DIV \mathring \Bvarphi)^\top p + V_{\Bvarphi} - \mathring \Bvarphi^\top \bB \right].
\end{split}
\end{align}
Here, we assume Einstein summation and $\tL = \mathcal{K}^\top \tA^{-1} \mathcal{K}$ is half of the discrete viscosity Laplacian (\cf \S\ref{sec: onsager principle}), where $A_{\lambda\gamma}$ is the area of the face \(\gamma\) for $\lambda=\gamma$ and zero otherwise.

Under any choice of \( V_{\Bvarphi} \) (assuming \(\bB = \boldsymbol 0\)), the variational integrator is unconditionally stable by design, as it preserves the system’s dissipative structure.
The dynamics follow a gradient flow of \( V_{\Bvarphi} \) in the metric space defined by the Stokes operator, ensuring a monotonic decrease in the Rayleighian: $\mathcal{R}_{(\epsilon)} \leq \mathcal{R}_{(0)}$. 
Therefore, the allowable time step size depends only on the solvability of the optimization in \eqref{eqn: discrete Rayleighian}, which can be efficiently handled using standard numerical optimization methods.
Here we use a simple gradient flow to solve this saddle-point optimization: $ \tH \dot \Bvarphi = -  \delta_{\Bvarphi} \mathcal{R}  =  -  2 \mu { \tL \mathring \Bvarphi }  - \epsilon ({\DIV^\top p } + {\delta_{\Bvarphi} V_{\Bvarphi}}  - {\bB}) $ and $h \dot p = \delta_{p} \mathcal R =   \DIV \mathring \Bvarphi$, where $\tH$ and $h$ provide a general metric for the gradient flow %
\footnote{We provide a MATLAB implementation available at \url{https://t.ly/vUxfh}.
The tensorial calculations in \eqref{eqn: discrete strain rate tensor} and \eqref{eqn: discrete Rayleighian} rely on \texttt{sptensor} \citep{bader2008efficient}. 
Remeshing is performed using SideFx Houdini.}. 
In \S\ref{sec: results}, we adopt the Helfrich Hamiltonian as $V$ and follow the discretization used in \citet{zhu2022mem3dg}.

As mentioned in \S\ref{sec: solvability}, equation \eqref{eqn: stokes flow} and its discrete version \eqref{eqn: discrete Rayleighian} have non-unique solution up to rigid body modes in $\Real^3$. 
To ensure that we obtain continuous dynamics during time evolution, we consistently project out rigid body modes.

\section{Results}\label{sec: results}
In this section, we validate the discrete model, analyze its convergence properties, and demonstrate its applicability to manifolds with intricate geometries and topologies. Numerical validation against analytical solutions is presented in \S\ref{sec: validation}, with further examples provided in \S\ref{sec: genus-6} and \S\ref{sec: bulk flow} to showcase its effectiveness.

\subsection{Validations}\label{sec: validation}
We validate the differential operators \(\mathcal{K}\) and \(\GRAD\) on a spheroid with semi-axes $a$ and $c$, parametrized by latitude $\beta \in [- \pi/2, \pi/2]$ and longitude $\theta \in (-\pi, \pi]$, using the realization $ \Bvarphi(\beta, \theta) = [ a \cos \beta \cos \theta, a \cos \beta \sin \theta, c \sin \beta ]$.

\begin{figure}
\centerline{\includegraphics[width= 0.95\textwidth]{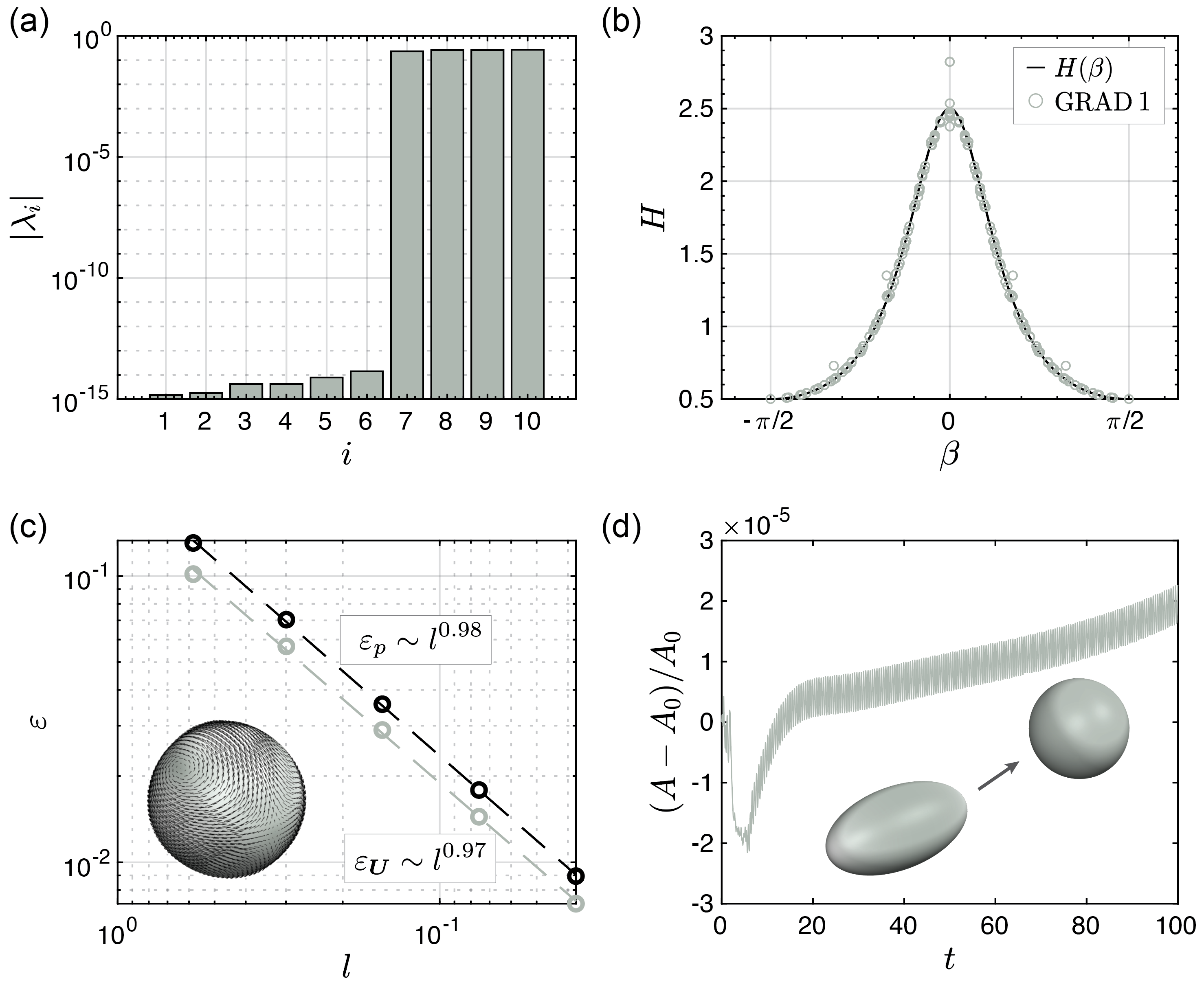}}%
  \caption{
Validation of our model and method: (a) Minimum eigenvalues $\lambda_{1\leq i \leq 10}$ of \(\tL \bU_i = \lambda_i \tA \bU_i\); (b) Comparison of numerical $\GRAD 1$ and analytical $H(\beta)$ evaluations of the mean curvature on a spheroid;  (c) Relative error $\varepsilon_{\bU}$ for the momentum equation and $\varepsilon_{p}$ for the continuity equation as functions of mean edge length $l$, with the inset showing the tangential forcing $\bb$ constructed from a spherical harmonic; (d) Relative error in total area $A$ compared to the initial area $A_0$ during Helfrich-Stokes relaxation, with insets showing the initial frame (prolate spheroid) and the final frame (sphere) of the relaxation. 
}
\label{fig:validation}
\end{figure}

In figure \ref{fig:validation}(a), we test the operator \(\mathcal K\) by computing the lowest ten eigenvalues of \( \tL = \mathcal {K}^\top \tA^{-1}\mathcal{K}\) with $\tL \bU_i = \lambda_i \tA \bU_i $ (\cf \eqref{eqn: discrete Rayleighian}).  
As predicted by the continuous theory, there are six vanishing eigenvalues, forming a six dimensional kernel of \(\mathcal{K}\) corresponding to the space of rigid body motions in \(\Real^3\).

To validate the operator \(\GRAD\) \eqref{eqn: divergence normal-tangent}, we check that the numerical evaluation of $ \tA^{-1}|\DIV^\top 1|$  aligns with the analytical mean curvature $H(\beta)$ of an oblate spheroid with $a = 1$ and $c = 0.5$, as shown in figure \ref{fig:validation}(b).
In fact, this evaluation of mean curvature is identical with the established cotan Laplacian discretization \citep{meyer2003discrete}.

We evaluate the convergence of our discretization of \eqref{eqn: stokes flow} by comparing it to an analytical solution on a unit sphere. 
The forcing term is analytically prescribed as \(\bB = \bb + b_n \bN =  \bN \times \grad \phi + b_n \bN \), where $\phi$ is a spherical harmonic satisfying $\Delta \phi \coloneqq \div \grad \phi = \lambda \phi$, \(\bN\) is the surface normal, and $b_n$ is a constant.
On the unit sphere, \(\bb\) is an eigenfunction of the viscosity Laplacian with eigenvalue $\lambda + 2$. 
The analytical solution \(\overline{\bU} = -\bb / (\lambda + 2)\) and $\overline{p} = -b_n / 2$ satisfies \eqref{eqn: stokes flow} ($\mu = 1$, $\delta_{\varphi} V_{\varphi} = 0$) exactly (\cf \eqref{eqn: stokes flow normal-tangent}).
Numerically, the relative residual of the momentum equation, \(\varepsilon_{\bU} = \|2 \tA^{-1} \tL \overline{\bU} - \tA^{-1} \DIV^\top \overline{p} - \bB \| / \|\bB \|\), and the residual of the continuity equation, \(\varepsilon_p = \|\tA^{-1} \DIV \overline \bU \|\), shown in figure~\ref{fig:validation}(c), decrease approximately linearly as the mesh is refined.

Lastly, figure \ref{fig:validation}(d) illustrates the total area of the fluid film during the Stokes relaxation ($\mu = 1$) from a prolate spheroid ($a = 1$, $c = 2$) to a sphere under Helfrich energy ($\kappa = 0.05 $), without mesh reparametrization.
Local incompressibility conserves the global area within a relative error of  $10^{-5}$ throughout the evolution.

\subsection{Relaxation of a genus-6 torus}\label{sec: genus-6}
\begin{figure}
\centerline{\includegraphics[width= \textwidth]{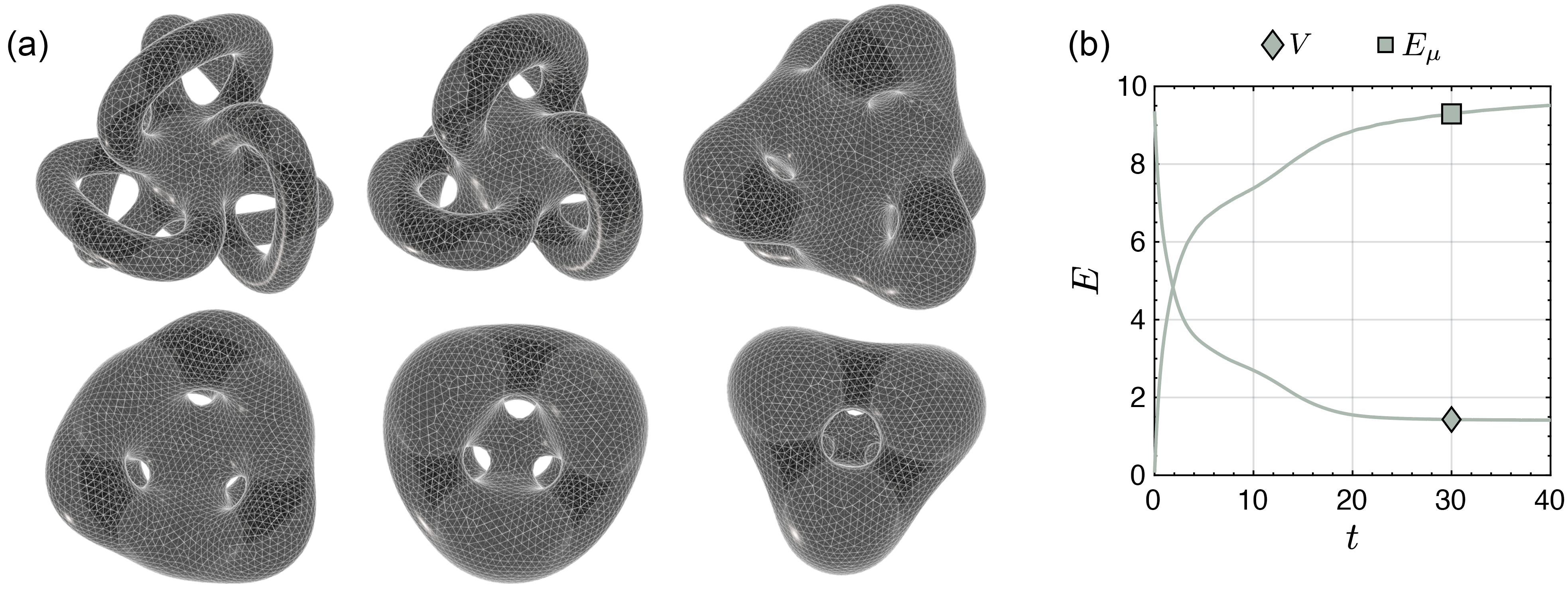}}%
  \caption{(a) Helfrich-Stokes relaxation of a genus-6 torus. 
 Snapshots are taken at $t = 0$, $2$, $8$, $14$, $20$, and $40$, from left to right, top to bottom. The animated simulation is available at \url{https://youtu.be/Llh0_N0hCPw} and in Movie 1 of the Supplementary Material.
  (b)~Elastic energy $V = \int \kappa H^2 \mathrm{d}A$ and cumulative dissipation $E_{\mu} = \int_0^t \llangle \tT | \tE \rrangle / 2|_\tau  \mathrm{d}\tau$, which theoretically sum to a constant value.} 
\label{fig:genus-6}
\end{figure}

We model the evolution of a high-genus, non-analytical lipid membrane ($\mu = 1$, $\kappa = 0.05$) based on the Helfrich-Stokes relaxation in figure \ref{fig:genus-6}(a).
In theory, the elastic Helfrich Hamiltonian should be monotonically dissipated through viscosity, $V + E_{\mu} = \int \kappa H^2 \mathrm{d}A + \int_0^t \llangle \tT | \tE \rrangle/ 2|_\tau \mathrm{d}\tau  = \mathrm{const}$. Although our numerical results in figure \ref{fig:genus-6}(b) do not exactly match this theoretical expectation, they show very good agreement.

\subsection{Deformation of a lipid membrane under bulk flow}\label{sec: bulk flow}
\begin{figure}
\centerline{\includegraphics[width= \textwidth]{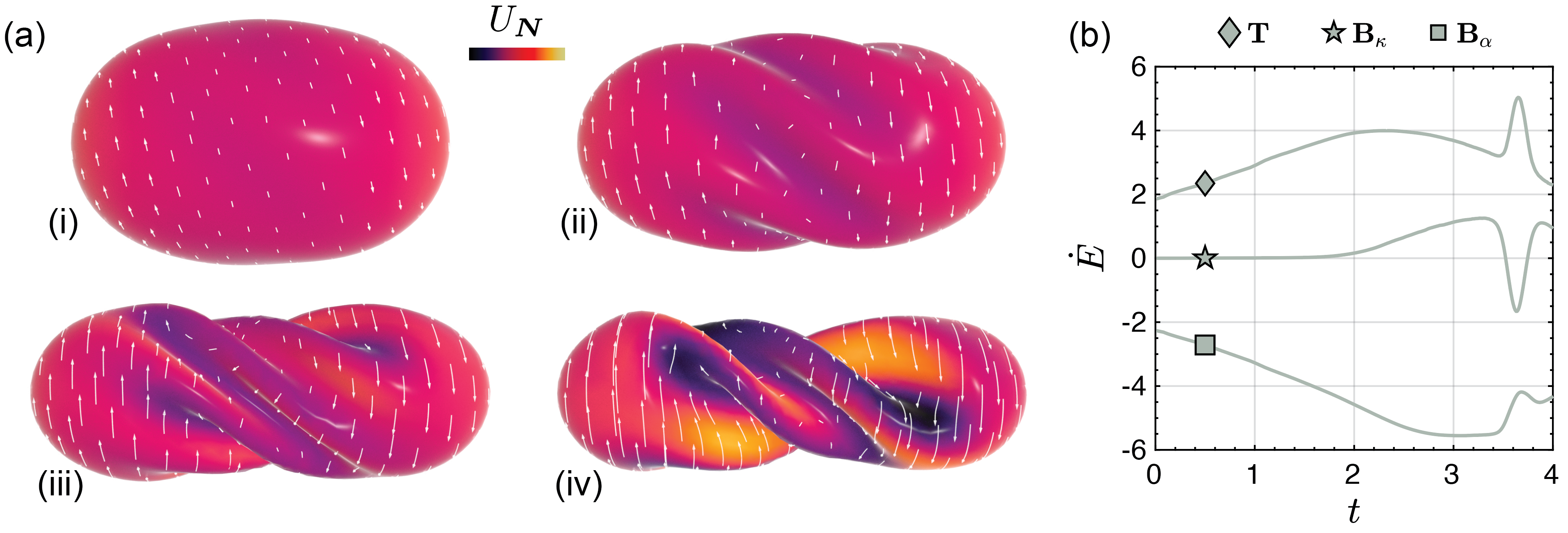}}%
  \caption{
  (a) Deformation of a lipid membrane under bulk flow, with the tangent velocity $\bu$ shown as streamlines and the normal velocity $U_{\bN}$ displayed in a colormap ranging from $-1$ to $1$ (\cf Appendix~\ref{sec: explicit formulation} for the tangent-normal decomposition).
  Snapshots (i)-(iv) are taken at $t = 1.5$, $2.5$, $3.4$, and $3.8$, respectively.
The animated simulation is available at \url{https://youtu.be/M0WsLihzRJk} and in Movie 2 of the Supplementary Material.
    (b) Energy dissipation rate $\dot E$ of the system due to viscosity $\dot{E}_{\mu} = \llangle \tT| \tE \rrangle / 2$, elasticity $\dot{E}_{\kappa} = \llangle \bB_\kappa | \bU \rrangle$, and bulk friction $\dot{E}_{\alpha} = \llangle \bB_\alpha | \bU \rrangle / 2$. }
\label{fig:bulk}
\end{figure}
We model the deformation of a spherical lipid vesicle ($\mu = 1$, $\kappa = 0.01$) under the friction ($\alpha = 0.5$, \cf \S\ref{sec: solvability}) from a constant bulk shear-extension flow defined by $\bU_0(\theta, r, z) = \bU_0^{\mathrm{shear}} + \bU_0^{\mathrm{ext}} = (5 r z) \hat \be_\theta + (- 0.5 r \hat \be_r + z \hat \be_z)$ in cylindrical coordinates.
As illustrated in figure~\ref{fig:bulk}(a), this bulk flow stretches and twists the initially spherical vesicle along the $z$-direction. 
A snapthrough-like instability occurs around $t = 3.7$ between frame (iii) and frame (iv) in figure \ref{fig:bulk}(a). 
While the bulk friction governs most of the dynamics, the Helfrich energy dominates during this buckling event, as indicated by the dip of the energy dissipation rate in figure \ref{fig:bulk}(b) around $t = 3.7$.

\section{Conclusion}
Using Onsager's variational principle, we derived and reformulated the evolving Stokes equations on a manifold.
We replaced the classic system of equations expressed in tangent-normal splitting with a simple, coordinate-free differential-geometric formulation. 
This approach directly leads to a straightforward discrete model and a numerical scheme to solve this long-standing problem in its full geometric generality.

Although the theoretical framework is not limited, the current minimal implementation is restricted to closed manifolds and employs first-order discretization.
Future work could incorporate boundary conditions and higher-order discretizations.
For applications, we expect the minimal system presented here to serve as a foundation for integrating more complex models for specific biophysical problems. 
These could include additional global volumetric/areal constraints, heterogeneity in material properties, in-plane anisotropy, and surface activity, as exemplified in \citet{zhu2022mem3dg, zhu2024active}.

\backsection[Acknowledgements]{The authors acknowledge funding from National Science Foundation Grant No. DMS-2153520 (D.S.) and National Science Foundation CAREER Award 2239062 (A.C.).
Additional support was provided by SideFX software.}

\backsection[Author ORCIDs]{\\ C. Zhu, https://orcid.org/0000-0003-1373-3492;\\ D. Saintillan, https://orcid.org/0000-0001-9948-708X;\\ A. Chern, https://orcid.org/0000-0002-9802-3619.}

\appendix

\section{Evolving Stokes equations through normal-tangent decomposition}\label{sec: explicit formulation}

\subsection{Differential operators for an evolving surface} \label{sec: differential operators for an evolving surface normal-tangent}
The second fundamental form (\ie curvature tensor), $\Second \in \Gamma(T^*M \otimes_{\operatorname{symm} }T^*M)$, is given by the derivative of the surface normal $\bN$: 
\begin{align}\label{eqn: second fundamental form}
    \Second \pair{\bv, \bw} =( \tF \bv )\bcdot({\bnabla}_{\tF \bw} \bN) \eqqcolon\langle \bv, \tS \bw \rangle,
\end{align}
where $\tS: \Gamma(TM) \rightarrow \Gamma(TM)$ is the shape operator. 

With the induced metric $\First$, there is an intrinsic Levi-Civita covariant derivative ${\bnabla}: \Gamma(TM) \rightarrow \Gamma(TM \otimes T^*M)$ on $M$.
The covariant derivative in $W$ used in \eqref{eqn: second fundamental form} and \eqref{eqn: deformation gradient dot} can be expressed in terms of the instrinsic covariant derivative and curvature tensor as:
\begin{align}
    {\bnabla}_{\tF \bw} \tF {\bv} = {\bnabla}_{\bw} \bv  - \Second \pair{\bw, \bv} \bN.
\end{align}

We can decompose the velocity into tangent and normal parts, $\bU = \tF {\bu} + U_N \bN$, where $\bu \in \Gamma (TM)$, $U_N \in C^\infty(W)$. 
Since \( \tF{\bu} \) effectively embeds the tangent vector \( \bu \) in $\Real^3$, we will abbreviate \( \tF {\bu} \) as \( \bu \) when unambiguous.
Given $\bv \in T_pM $, \eqref{eqn: deformation gradient dot} can be expressed as ${\bnabla}_{\tF \bv} \bU   = { {\bnabla}_{\bv} \bu + U_N \tS \bv} +  \bN  \left( \langle \grad U_N, \bv \rangle - \Second \pair {\bu, \bv} \right)$.
Combining with \eqref{eqn: strain rate} and \eqref{eqn: evolving killing}, we have $ 2 (\mathcal{K} \bU) \pair{\bv, \bw}  =   \langle {\bnabla}_{\bv} \bu, \bw \rangle + \langle {\bnabla}_{\bw} \bu , \bv \rangle + 2 U_N \Second \pair{\bv, \bw}$, or 
\begin{align}\label{eqn: killing normal-tangent}
 2 \mathcal{K} \bU= {\bnabla} \bu + ({\bnabla} \bu)^\top + 2 U_N \Second \eqqcolon 2 \mathscr{K} \bu + 2 U_N \Second,
\end{align}
where $\mathscr{K}: \Gamma(TM) \rightarrow \Gamma(T^*M \otimes_{\symm} T^*M)$ denotes the intrinsic Killing operator. 
Combining \eqref{eqn: evolving divergence},  \eqref{eqn: evolving gradient} and \eqref{eqn: killing normal-tangent}, the divergence and gradient can be expressed as
\begin{align}\label{eqn: divergence normal-tangent}
    \operatorname{DIV} \bU = \div\bu + 2H U_N , \quad \GRAD(p) = \grad p - 2 p H \bN,
\end{align}
where $\div = \tr (\circ \mathscr{K})$ is the intrinsic divergence, $\grad = -\div^*$ is the intrinsic gradient, and $H = \tr \Second /2$ is the mean curvature of the surface. 
From \eqref{eqn: evolving covariant divergence} and \eqref{eqn: killing normal-tangent}, we get:
\begin{align}\label{eqn: covariant divergence normal-tangent}
    \DIV^{\bnabla} \tT = \div^{\bnabla} \tT - \langle \Second | \tT \rangle ~ \bN, 
\end{align}
where $\div^{\bnabla} = -\mathscr K^*: \Gamma(TM \otimes TM) \rightarrow \Gamma(T^*M)$ is the intrinsic covariant divergence.

\subsection{Evolving Stokes equations}\label{sec: evolving stokes equations normal-tangent}
The viscosity Laplacian for an evolving surface can be explicitly decomposed as: 
\begin{align}\label{eqn: viscosity laplacian normal-tangent}
\begin{split}
  2 \DIV^{\bnabla} \mathcal{K} \bU =&\, 2 \left[ \div^{\bnabla} \mathscr{K} \bu +  \div^{\bnabla} (U_N\Second) - \left ( \langle \Second , {\mathscr{K} \bu} \rangle  + U_N |\Second|^2 \right) \bN \right] \\ 
   =&\,  ( \BDelta + K + \grad \circ \div)\bu + 2 \tS (\grad U_N) + 4 U_N \grad H 
   \\
   &%
   -2  \left [ \langle {\bnabla} \bu, \Second \rangle +  U_N (4 H^2 - 2K) \right] \bN,
\end{split}
\end{align}
where $\BDelta = \div^{\bnabla} {\bnabla}: \Gamma(TM) \rightarrow \Gamma(T^*M)$ is the intrinsic Bochner Laplacian, and $K = \det \Second$ is the Gaussian curvature. 

To summarize, by substituting \eqref{eqn: killing normal-tangent}-\eqref{eqn: viscosity laplacian normal-tangent} 
into \eqref{eqn: stokes flow}, we get the system of evolving Stokes equations subject to external force $\bB = \bb + b_n \bN$ and bending force $\bB_\kappa =  \kappa \bN \left( \Delta H +   2 H ( H^2 - K) \right)$:
\begin{align}\label{eqn: stokes flow normal-tangent}
\begin{split}
    \mu \left [ (\BDelta + K)\bu  + 2 \tQ (\grad U_N) + 2 U_N \grad H \right ] - \grad p  + \bb &= 0, \\
    -2 \mu \left [  \langle {\bnabla} \bu , \Second \rangle  + U_N (4H^2 - 2K)  \right] + 2 p H + b_n +  \kappa \left( \Delta H +   2 H ( H^2 - K) \right) &= 0, \\
    \div \bu + 2H U_N &= 0.\\
\end{split}
\end{align}
Note that the Hopf differential $\tQ = \tS - H \First$ is the traceless part of the shape operator $\tS$ that represents the deviatoric curvature. 
Formulation \eqref{eqn: stokes flow normal-tangent} agrees with earlier results by \citet{scriven1960dynamics} and \citet{arroyo_relaxation_2009}.

\bibliographystyle{jfm}
\bibliography{ref}

\end{document}